\documentclass[preprint]{aastex}
\slugcomment{To appear in Ap.J.\ Lett.}
\shorttitle{Hundred Thousand Degree Gas in Virgo}
\shortauthors{Sparks et al.}
\begin{document}
\title{Hundred Thousand Degree Gas in the Virgo Cluster of Galaxies}
\author{W.B.\ Sparks,\altaffilmark{1} J.E.\ Pringle,\altaffilmark{1,2} 
R.F.\ Carswell,\altaffilmark{2} M.~Donahue,\altaffilmark{3} 
R.~Martin,\altaffilmark{1} M.~Voit,\altaffilmark{3}\\ M.~Cracraft,\altaffilmark{1} 
N.~Manset,\altaffilmark{4} \& J.H.\ Hough\/\altaffilmark{5}}
\affil{1. Space Telescope Science Institute, 3700 San Martin Drive, Baltimore, MD 21218}
\affil{2. Institute of Astronomy, University of Cambridge, Madingley Road, Cambridge, CB3~0HA, UK}
\affil{3. Dept of Physics and Astronomy, Michigan State University, East Lansing, MI~48824-2320}
\affil{4. CFHT, 65-1238 Mamalahoa Hwy., Kamuela, HI 96743}
\affil{5. University of Hertfordshire, College Lane, Hatfield, AL10 9AB, UK}
\email{sparks@stsci.edu}

\begin{abstract}
The physical relationship between low-excitation gas filaments at $\sim$10$^4$~K, seen in optical line emission, and diffuse X-ray emitting coronal gas at $\sim$10$^7$~K in the centers of many galaxy clusters is not understood.  It is unclear whether the $\sim$10$^4$~K filaments have cooled and condensed from the ambient hot ($\sim$10$^7$~K) medium or have some other origin such as the infall of cold gas in a merger, or the disturbance of an internal cool reservoir of gas by nuclear activity. Observations of gas at intermediate temperatures ($\sim$10$^5$--$10^6$~K) can potentially reveal whether the central massive galaxies are gaining cool gas through condensation or losing it through conductive evaporation and hence identify plausible scenarios for transport processes in galaxy cluster gas. Here we present spectroscopic detection of $\sim$10$^5$~K gas spatially associated with the $H\alpha$ filaments in a central cluster galaxy, M87 in the Virgo Cluster.  The measured emission-line fluxes from triply ionized carbon (C\,IV~1549~\AA) and singly ionized helium (He\,II~1640~\AA) are consistent with a model in which thermal conduction determines the interaction between hot and cold phases.
\end{abstract}

\section{INTRODUCTION}
Galaxies generally grow by accreting smaller galaxies and intergalactic gas clouds, but as a galaxy's mass increases, the accreting gas is heated to progressively greater temperatures, inhibiting its ability to cool and form stars. Nevertheless, some of the most massive galaxies in the universe, found at the centers of galaxy clusters, still manage to harbor cool gas capable of fueling star formation
\citep{2001MNRAS.328..762E}. The most obvious manifestation of this cool medium is optical line emission from filamentary gas at $\sim$10$^4$~K. It is not yet clear whether these filaments have cooled and condensed from the ambient hot 
($\sim$10$^7$~K) medium or have some other origin
\citep{1977MNRAS.180..479F, 1989ApJ...345..153S,2001MNRAS.328..762E}.
Observations of gas at intermediate temperatures ($\sim$10$^5$--$10^6$~K) are crucial for solving this puzzle, because they can potentially reveal whether these massive galaxies are gaining cool gas through condensation or losing it through conductive evaporation.  We therefore developed simple conductive models to estimate the approximate strength of line emission from intermediate temperature gas, and carried out observations to test those predictions. Here we present spectroscopic detection of $\sim$10$^5$~K gas spatially associated with the filaments in a central cluster galaxy.  The measured emission-line fluxes from triply ionized carbon (C\,IV~1549~\AA) and singly ionized helium (He\,II~1640~\AA) are consistent with a model in which thermal conduction determines the interaction between hot and cold phases.

Central cluster galaxies with optically emitting filaments tend to reside in galaxy clusters with the greatest central surface brightness in X-rays,  cool-core clusters.  The cooler core temperatures may arise because the central gas is capable of radiating away its thermal energy in less than a few billion years, but that is not the only interpretation. An alternative hypothesis suggests that the optically emitting gas at 
$\sim$10$^4$~K may be the cause of the cool-core phenomenon, if electron diffusion effectively conducts thermal energy from the hot gas into the cooler filaments
\citep{1989ApJ...345..153S}. Conductive heat transfer from the X-ray gas to the much cooler optically emitting gas and its associated dust can be similar in magnitude to the total emitted radiation from the cool phase, suggesting that conduction might be the energy source for the observed optical and infra-red emission and that this transfer of energy may lower the temperature of the neighboring X-ray emitting gas, thereby raising its X-ray emissivity if pressure balance is maintained. However, despite decades of study, there is still no consensus on the origin and excitation mechanisms of the filaments in cool-core clusters and their relevance to the fuelling of star formation in massive galaxies \citep{1997ApJ...486..242V, 2004cgpc.symp..143D, 2005MNRAS.363..216C, 2007MNRAS.380...33H, 2009MNRAS.392.1475F,
2010ApJ...721.1262M}.

Progress on these questions is likely to come from investigations focusing on the gas at temperatures in between those of the hot X-ray emitting gas and the optical line emitting gas at $\sim$10$^4$~K. If there were no thermal communication between the hot and cold phases, there ought not be any intermediate temperature gas; but if there is a permeable interface between the two, and consequently energy and possibly mass flux, then intermediate temperature gas will be present in quantities that depend on the nature of the physical process \citep{1992ApJ...399...66S, 2009MNRAS.392.1475F, 2010MNRAS.407.2063W, 2011MNRAS.417..172F}.

M87 is the central galaxy of the Virgo galaxy cluster at a distance of 16.7 Mpc
\citep{2009ApJ...694..556B}, exhibiting a well-known central X-ray luminosity peak, characteristic of cool-core clusters, along with a low-excitation optical filament system \citep{1979ApJS...41..147F}. The proximity of M87 and its high apparent X-ray luminosity in the sky ``makes it a perfect target for studying the processes that are important in all clusters of galaxies, but in much greater detail than possible elsewhere'' and ``a famous cool core cluster'' \citep{werner2006}. Nevertheless, it is the case that the Virgo cluster is an intrinsically low power cool core cluster, and more extreme clusters exhibit a more complex variety of phenomena such as star formation and powerful AGN. The simpler physics of Virgo offers a better prospect for reaching an understanding, while its applicability to the more extreme versions is not known.

In addition to low excitation optical filaments, a recent FUV HST image of M87 showed evidence for much hotter $10^5$~K gas associated with the optical filaments
\citep{2009ApJ...704L..20S}. The FUV brightness of these filaments was consistent with model predictions of C\,IV 1549~\AA\  emission from intermediate-temperature ($\sim10^5$~K) gas in a conductive interface between the optical filaments and the hot X-ray emitting gas.  However, a re-analysis of the image concluded that the filamentary FUV emission could be continuum emission from hot stars, which would also be capable of exciting optical line emission from the cooler gas
\citep{2011MNRAS.414.2309O}.  Here we present new deep FUV images and spectroscopy that  confirm the emission as intermediate temperature gas.

\section{OBSERVATIONS}

The earlier short exposure FUV image was obtained with the Advanced Camera for Surveys Solar Blind Channel (ACS/SBC) onboard the Hubble Space Telescope (HST) using the 150~nm long-pass filter, which transmits FUV emission from 150--200~nm \citep{2009ApJ...704L..20S}. We carried out additional FUV HST observations of the M87 filament system in Cycle~18, in which we obtained four orbits (11,536~sec) imaging with the ACS/SBC shown in Fig.~1 and two orbits (5025~sec) using the FUV Cosmic Origins Spectrograph (COS) spectrograph with the low resolution G140L grating. The COS with the G140L grating acquires data from 
90~nm to 200~nm using the circular  2.5~arcsec primary science aperture. We acquired four spectra each obtained in TIME-TAG mode using a different FP-POS setting to eliminate detector artifacts. The spectra were reduced using standard pipeline processing software and flux calibration, and to provide additional robustness against instrumental or software artifacts, we also worked with a median combination of the four individually extracted spectra for each observation. This median COS spectrum is shown in Fig.~2.

The C\,IV~1549~\AA\  emission line is clearly visible, along with emission from 
He\,II~1640~\AA. We measure a C\,IV flux of $(5.54\pm 0.36)\times10^{-16}$~erg~s$^{-1}$~cm$^{-2}$ within a 2.5~arcsec circular aperture, and a flux of $(2.16\pm 0.48)\times10^{-16}$~erg~s$^{-1}$~cm$^{-2}$ for the He\,II line. These fluxes sum to $7.70\times10^{-16}$~erg~s$^{-1}$~cm$^{-2}$, which is very close to the total flux of $7.94\times10^{-16}$~erg~s$^{-1}$~cm$^{-2}$ inferred from the original image by placing a circular 2.5~arcsec aperture at the location of the spectroscopic observation. The agreement may be slightly fortuitous given the quoted  standard errors on the mean from measuring the four spectra individually. 
 
\section{DISCUSSION}

The emissivity of C\,IV has a  well-defined maximum at temperature  $\approx$10$^5$~K. He\,II also peaks in emissivity at $\sim$10$^5$~K, though has a shallower decline in emissivity than C\,IV as the temperature increases above $10^5$~K. This spectroscopic data therefore unequivocally confirms the presence of FUV line emission in this region of M87, and establishes that filamentary gas is present in the Virgo Cluster at $\sim$10$^5$~K.

There is no evidence in these observations for any additional FUV continuum emission associated with the filament system. The spectroscopically measured FUV surface brightness of the continuum in the COS aperture ($4.3\times10^{-18}$~erg~s$^{-1}$~cm$^{-2}$~\AA$^{-1}$~arcsec$^{-2}$) is similar to the FUV brightness of the underlying galaxy ($5.4\times10^{-18}$~erg~s$^{-1}$~cm$^{-2}$~\AA$^{-1}$~arcsec$^{-2}$) measured from the earlier image by modeling the galaxy as a set of smooth ellipses. Hot young stars would have been seen individually in the earlier image, while the short timescale of residence for stars within such filaments excludes hot old stars \citep{2009ApJ...704L..20S}. The new empirical data reinforce these arguments: we still do not see individual stars using the new deep image, and the FUV spectra reveal no excess continuum above the elliptical body of the galaxy.

Even though we have shown FUV continuum emission from hot stars is insignificant in this part of M87, it is possible that it may be important in other central cluster galaxies.  It has been shown that FUV continuum emission is present in the centers of some galaxy clusters \citep{1995ApJ...443...77M, 2011ApJ...734...95M}, which is not surprising since central star formation has previously been found in a number of cool-core galaxy clusters \citep{1972ApJ...176....1G,1998MNRAS.298..977C}. Yet, it still remains unclear whether the FUV light from newly formed stars dominates the ionization and excitation of the  filaments in more massive clusters in general. 

Conduction models for radiating filaments should properly take account of time-dependent flows---either a condensing flow onto the filament or an evaporating flow away from the filament---depending on circumstances. For simplicity we consider here only an intermediate model: inflow of heat due to thermal conduction from the ambient medium is exactly balanced by radiation and there is no velocity flow \citep{2004MNRAS.349.1509N}. This model contrasts with those of \citet{1989MNRAS.237.1147B} who consider line radiation from filaments which are either evaporating or condensing at steady fixed rates, but who neglect to take account of the ionization energies released/absorbed by the cooling/heating atoms as they pass through the flow. The models we computed assumed cylindrical filaments with a fixed inner temperature of $\log T_{\rm in} = 3.75$ ($T_{\rm in} = 5623$~K ) and $dT/dR = 0$ at an inner radius $R_{\rm in}$ of 10~pc, so that no heat was conducted into the cool body of the filaments, and full Spitzer conductivity. With conductivity $\propto T^{7/2}$ the conductivity is very low near the inner boundary so our results are insensitive the the exact temperature chosen. Their emission-line properties were calculated using the code `CLOUDY' \citep{1998PASP..110..761F} (http://www.nublado.org) with Solar elemental abundances. Our treatment of conduction allows for the possibility that the heat flux might become saturated when the electron mean-free-path becomes longer than the scale-length for temperature variation and when the conducting electrons are able to stream freely at the (ion) sound speed \citep{1977ApJ...211..135C}. However, for the model parameters adopted here classical thermal conduction obtained at all times. For a pressure of $2.5\times10^{-9}$~dynes~cm$^{-2}$ in M87 \citep{2004ApJ...607..294S}, the derived 
C\,IV 1549~\AA\  flux for a filament length corresponding to the 2.5~arcsec COS aperture is $3.83\times10^{-16}$~erg~s$^{-1}$~cm$^{-2}$, and He\,II~1640~\AA\  similar at $\approx$$3.8\times10^{-16}$~erg~s$^{-1}$~cm$^{-2}$,  both within a factor two of the observations and well within the uncertainties. This casts doubt on the conclusion by  \citet{1989MNRAS.237.1147B} that their ``...results show that the
luminosities of the optical and UV emission line emission from warm gas in cooling flows cannot plausibly be due to conductive interfaces.''

We stress that we have not attempted to develop a fully self-consistent model. We just worked through a simple, example conduction model and find that in fact, it is consistent with the data, even given significant uncertainties, at least a factor of a few, on the geometry and time evolution of the system. The model cannot predict the optical line strengths because it is terminated at the filament boundary and different physics are likely to be relevant inside the filament. However, in \citet{2004ApJ...607..294S} we showed that energetically, the surface brightness of the $H\alpha$ line emission is consistent with heat available through  conduction for a variety of models of the coronal density and temperature.

By contrast, \citet{2011MNRAS.417..172F} advocate a scenario for similar filaments in NGC~1275 where there are two co-existing phases, the hot and cold phases, ``which do not interact via classical thermal conduction,'' and an unquantified amount of intermediate temperature gas. Likewise, it is well-known that in the condensation or cooling-flow scenario, the cooling is supposedly distributed over a much wider region than the cool filaments, and that the power of the cool filaments far exceeds expectation for a single recombination. Without predictions for the amount of intermediate temperature gas in their scenario, it is difficult to understand whether or not there is a version of that model consistent with the new data. For example, if one were to naively assume each C\,IV photon corresponded to a recombining carbon atom, the implied mass deposition rate would be many thousands of Solar masses per year, a similar situation to that found in $H\alpha$. If the C\,IV is from gas cooling and condensing onto the filament, then one would expect self-irradiation to boost the $H\alpha / $C\,IV ratio to greater levels than is observed \citep{1994ApJS...95...87V}. The cooling rate in that model would be $\approx$5~M$_{\Sun}$/yr, however in addition to the higher predicted overall $H\alpha/$C\,IV ratio, there ought to be large-scale variations in $H\alpha/$C\,IV owing to differences in the geometrical relationship of cooling gas with respect to irradiated gas, which is not observed \citep{2009ApJ...704L..20S}. Rather, the approximate constancy of that ratio argues for a locally self-regulating process such as conduction. Also the filaments are dusty \citep{1993ApJ...413..531S}, which would not be expected if they have condensed from a hot coronal phase.

If thermal conduction is operating efficiently, it can potentially reverse the expected sense of mass flow in the central galaxy, causing cool gas clouds to evaporate instead of condensing and forming stars.  In that case, the cool optically-emitting clouds would not come from condensation of the hot ambient medium but would have some other source.  Possible sources of cool gas include ram-pressure stripping of gas-rich galaxies passing through the cluster core \citep{1993ApJ...414L..17D, 2011AJ....141..164A}
as well as normal stellar mass loss from the stars of the central galaxy \citep{2011ApJ...738L..24V}.  Either of those mass sources would more naturally explain why the cool gas in cluster cores contains small dust particles, \citep{1989ApJ...345..153S, 2011ApJ...732...40D} which are rapidly destroyed by sputtering in the hot ambient medium \citep{2010A&A...518L..50C}. From \citet{2007ApJ...659L.115C}, approximately $10^{10}$~M$_{\Sun}$ of gas is currently being stripped by Virgo cluster spirals, in timescales ranging from 100~Myr to 1~Gyr, implying a deposition rate of order 10--100~M$_{\Sun}$/yr. The classical evaporation timescale for a 10~pc sphere in pressure equilibrium is $\sim$10$^8$~yr \citep{1977ApJ...211..135C}, and the total mass of the filament system is $\sim$10$^7$~M$_{\Sun}$ \citep{1993ApJ...413..531S} so if the filaments are evaporating at this classical non-radiative rate, gas would need to be resupplied at $\sim$0.1~M$_{\Sun}$/yr, a small, and therefore plausible, fraction of the amount currently being stripped.

If conduction does indeed evaporate cool gas in large elliptical galaxies, then it may be crucial to the suppression of star formation within them.  Star formation is not present in the central galaxies of galaxy clusters in which conduction can plausibly channel enough heat inward to resupply the energy radiated by hot gas in the cluster core
\citep{2008ApJ...681L...5V,2011ApJ...740...28V}.  In those systems, conduction can quickly eradicate cool gas clouds and suppress star formation on its own.  However, in cool-core systems like the Virgo cluster, whose denser cores promote more rapid radiative energy loss, cool gas clouds can persist against conductive evaporation for a longer time and can go on to form stars if radiative cooling successfully preserves them.  
It is widely believed that heat input triggered by accretion of cooling gas onto a central supermassive black hole globally limits the star formation rate in such systems
\citep{2007ARA&A..45..117M}. And indeed, the Virgo cluster contains just such a black hole, which is currently in the midst of an outburst.  However, the outburst deposits heat into the hot ambient medium, not the cool filaments, meaning that conduction, which naturally channels thermal energy to the coolest regions, may still be necessary to energize the observed line emission and prevent star formation. 

\section{CONCLUSIONS}

The  observations presented here, FUV spectroscopy and deep imaging with HST,  show unequivocally that there is filamentary intermediate temperature gas at $\sim$10$^5$~K associated with the low-excitation optical filaments of M87 in the Virgo Cluster. This is strong evidence that there is an intermediate-temperature interface between the cool optically-emitting filaments in M87 and the surrounding hot gas, with a temperature distribution suggesting that an energy transport process, plausibly electron thermal conduction, is linking the two phases. This in turn has implications for our understanding of the physics and evolution of the gas in and around the massive central galaxies in galaxy clusters, as conduction and other similar energy transport processes fundamentally affect its thermal stability, cooling timescale, and energy budget.  Future observations of ultraviolet emission lines from intermediate temperature gas offer the prospect of providing uniquely powerful insight into the fundamental physics governing the energetics and stability of the gas in galaxy clusters, including assessing the importance of thermal conduction in cluster cores and its role in the suppression of star formation in the universe's largest galaxies, as well as other competing physical processes.

\acknowledgments
These data were obtained using the Hubble Space Telescope which is operated by STScI/AURA under grant NAS5-26555. We acknowledge support from grant HST/GO-12271. JEP thanks the Distinguished Visitor Program at STScI for its continued hospitality. MD \& MV acknowledge partial support of NASA LTSA
award NNG-05GC82G.

{\it Facilities:} \facility{HST (COS)}, \facility{HST (ACS)}.

\bibliographystyle{apj}                       



\begin{thebibliography}{35}

\expandafter\ifx\csname natexlab\endcsname\relax\def\natexlab#1{#1}\fi

\bibitem[{{Abramson} {et~al.}(2011){Abramson}, {Kenney}, {Crowl}, {Chung}, {van
  Gorkom}, {Vollmer}, \& {Schiminovich}}]{2011AJ....141..164A}
{Abramson}, A., {Kenney}, J.~D.~P., {Crowl}, H.~H., {et~al.} 2011, \aj, 141,
  164

\bibitem[{{Blakeslee} {et~al.}(2009){Blakeslee}, {Jord{\'a}n}, {Mei},
  {C{\^o}t{\'e}}, {Ferrarese}, {Infante}, {Peng}, {Tonry}, \&
  {West}}]{2009ApJ...694..556B}
{Blakeslee}, J.~P., {Jord{\'a}n}, A., {Mei}, S., {et~al.} 2009, \apj, 694, 556

\bibitem[{{Boehringer} \& {Fabian}(1989)}]{1989MNRAS.237.1147B}
{Boehringer}, H., \& {Fabian}, A.~C. 1989, \mnras, 237, 1147

\bibitem[{{Cardiel} {et~al.}(1998){Cardiel}, {Gorgas}, \&
  {Aragon-Salamanca}}]{1998MNRAS.298..977C}
{Cardiel}, N., {Gorgas}, J., \& {Aragon-Salamanca}, A. 1998, \mnras, 298, 977

\bibitem[{{Chung} {et~al.}(2007){Chung}, {van Gorkum}, \&
  {Kenney}}]{2007ApJ...659L.115C}
{Chung}, A., {van Gorkum}, J.~H., \& {Kenney}, J.~D.~P. 2007, \apjl, 659, L115

\bibitem[{{Clemens} {et~al.}(2010){Clemens}, {Jones}, {Bressan}, {Baes},
  {Bendo}, {Bianchi}, {Bomans}, {Boselli}, {Corbelli}, {Cortese}, {Dariush},
  {Davies}, {de Looze}, {di Serego Alighieri}, {Fadda}, {Fritz},
  {Garcia-Appadoo}, {Gavazzi}, {Giovanardi}, {Grossi}, {Hughes}, {Hunt},
  {Madden}, {Pierini}, {Pohlen}, {Sabatini}, {Smith}, {Verstappen}, {Vlahakis},
  {Xilouris}, \& {Zibetti}}]{2010A&A...518L..50C}
{Clemens}, M.~S., {Jones}, A.~P., {Bressan}, A., {et~al.} 2010, \aap, 518, L50

\bibitem[{{Cowie} \& {McKee}(1977)}]{1977ApJ...211..135C}
{Cowie}, L.~L., \& {McKee}, C.~F. 1977, \apj, 211, 135

\bibitem[{{Crawford} {et~al.}(2005){Crawford}, {Hatch}, {Fabian}, \&
  {Sanders}}]{2005MNRAS.363..216C}
{Crawford}, C.~S., {Hatch}, N.~A., {Fabian}, A.~C., \& {Sanders}, J.~S. 2005,
  \mnras, 363, 216

\bibitem[{{Donahue} {et~al.}(2011){Donahue}, {de Messi{\`e}res}, {O'Connell},
  {Voit}, {Hoffer}, {McNamara}, \& {Nulsen}}]{2011ApJ...732...40D}
{Donahue}, M., {de Messi{\`e}res}, G.~E., {O'Connell}, R.~W., {et~al.} 2011,
  \apj, 732, 40

\bibitem[{{Donahue} \& {Voit}(1993)}]{1993ApJ...414L..17D}
{Donahue}, M., \& {Voit}, G.~M. 1993, \apjl, 414, L17

\bibitem[{{Donahue} \& {Voit}(2004)}]{2004cgpc.symp..143D}
\underline{~~~~~~~}. 2004, Clusters of Galaxies: Probes of Cosmological Structure and Galaxy  Evolution, 143

\bibitem[{{Edge}(2001)}]{2001MNRAS.328..762E}
{Edge}, A.~C. 2001, \mnras, 328, 762

\bibitem[{{Fabian} \& {Nulsen}(1977)}]{1977MNRAS.180..479F}
{Fabian}, A.~C., \& {Nulsen}, P.~E.~J. 1977, \mnras, 180, 479

\bibitem[{{Fabian} {et~al.}(2011){Fabian}, {Sanders}, {Williams}, {Lazarian},
  {Ferland}, \& {Johnstone}}]{2011MNRAS.417..172F}
{Fabian}, A.~C., {Sanders}, J.~S., {Williams}, R.~J.~R., {et~al.} 2011, \mnras,   417, 172

\bibitem[{{Ferland} {et~al.}(2009){Ferland}, {Fabian}, {Hatch}, {Johnstone},
  {Porter}, {van Hoof}, \& {Williams}}]{2009MNRAS.392.1475F}
{Ferland}, G.~J., {Fabian}, A.~C., {Hatch}, N.~A., {et~al.} 2009, \mnras, 392, 1475

\bibitem[{{Ferland} {et~al.}(1998){Ferland}, {Korista}, {Verner}, {Ferguson},
  {Kingdon}, \& {Verner}}]{1998PASP..110..761F}
{Ferland}, G.~J., {Korista}, K.~T., {Verner}, D.~A., {et~al.} 1998, \pasp, 110,  761

\bibitem[{{Ford} \& {Butcher}(1979)}]{1979ApJS...41..147F}
{Ford}, H.~C., \& {Butcher}, H. 1979, \apjs, 41, 147

\bibitem[{{Gunn} \& {Gott}(1972)}]{1972ApJ...176....1G}
{Gunn}, J.~E., \& {Gott}, III, J.~R. 1972, \apj, 176, 1

\bibitem[{{Hatch} {et~al.}(2007){Hatch}, {Crawford}, \&
  {Fabian}}]{2007MNRAS.380...33H}
{Hatch}, N.~A., {Crawford}, C.~S., \& {Fabian}, A.~C. 2007, \mnras, 380, 33

\bibitem[{{McDonald} {et~al.}(2010){McDonald}, {Veilleux}, {Rupke}, \&
  {Mushotzky}}]{2010ApJ...721.1262M}
{McDonald}, M., {Veilleux}, S., {Rupke}, D.~S.~N., \& {Mushotzky}, R. 2010,
  \apj, 721, 1262
  
  \bibitem[{{McDonald} {et~al.}(2011){McDonald}, {Veilleux}, {Rupke},
  {Mushotzky}, \& {Reynolds}}]{2011ApJ...734...95M}
{McDonald}, M., {Veilleux}, S., {Rupke}, D.~S.~N., {Mushotzky}, R., \&
  {Reynolds}, C. 2011, \apj, 734, 95

\bibitem[{{McNamara}(1995)}]{1995ApJ...443...77M}
{McNamara}, B.~R. 1995, \apj, 443, 77

\bibitem[{{McNamara} \& {Nulsen}(2007)}]{2007ARA&A..45..117M}
{McNamara}, B.~R., \& {Nulsen}, P.~E.~J. 2007, \araa, 45, 117

\bibitem[{{Nipoti} \& {Binney}(2004)}]{2004MNRAS.349.1509N}
{Nipoti}, C., \& {Binney}, J. 2004, \mnras, 349, 1509

\bibitem[{{Oonk} {et~al.}(2011){Oonk}, {Hatch}, {Jaffe}, {Bremer}, \& {van
  Weeren}}]{2011MNRAS.414.2309O}
{Oonk}, J.~B.~R., {Hatch}, N.~A., {Jaffe}, W., {Bremer}, M.~N., \& {van
  Weeren}, R.~J. 2011, \mnras, 414, 2309

\bibitem[{{Sparks}(1992)}]{1992ApJ...399...66S}
{Sparks}, W.~B. 1992, \apj, 399, 66

\bibitem[{{Sparks} {et~al.}(1993){Sparks}, {Ford}, \&
  {Kinney}}]{1993ApJ...413..531S}
{Sparks}, W.~B., {Ford}, H.~C., \& {Kinney}, A.~L. 1993, \apj, 413, 531

\bibitem[{{Sparks} {et~al.}(2004){Sparks}, {Donahue}, {Jord{\'a}n},
  {Ferrarese}, \& {C{\^o}t{\'e}}}]{2004ApJ...607..294S}
{Sparks}, W.~B., {Donahue}, M., {Jord{\'a}n}, A., {Ferrarese}, L., \&
  {C{\^o}t{\'e}}, P. 2004, \apj, 607, 294

\bibitem[{{Sparks} {et~al.}(1989){Sparks}, {Macchetto}, \&
  {Golombek}}]{1989ApJ...345..153S}
{Sparks}, W.~B., {Macchetto}, F., \& {Golombek}, D. 1989, \apj, 345, 153

\bibitem[{{Sparks} {et~al.}(2009){Sparks}, {Pringle}, {Donahue}, {Carswell},
  {Voit}, {Cracraft}, \& {Martin}}]{2009ApJ...704L..20S}
{Sparks}, W.~B., {Pringle}, J.~E., {Donahue}, M., {et~al.} 2009, \apjl, 704,  L20

\bibitem[{{Voit}(2011)}]{2011ApJ...740...28V}
{Voit}, G.~M. 2011, \apj, 740, 28

\bibitem[{{Voit} {et~al.}(2008){Voit}, {Cavagnolo}, {Donahue}, {Rafferty},
  {McNamara}, \& {Nulsen}}]{2008ApJ...681L...5V}
{Voit}, G.~M., {Cavagnolo}, K.~W., {Donahue}, M., {et~al.} 2008, \apjl, 681, L5

\bibitem[{{Voit} \& {Donahue}(1997)}]{1997ApJ...486..242V}
{Voit}, G.~M., \& {Donahue}, M. 1997, \apj, 486, 242

\bibitem[{{Voit} \& {Donahue}(2011)}]{2011ApJ...738L..24V}
\underline{~~~~~~~~}. 2011, \apjl, 738, L24

\bibitem[{{Voit} {et~al.}(1994){Voit}, {Donahue}, \&
  {Slavin}}]{1994ApJS...95...87V}
{Voit}, G.~M., {Donahue}, M., \& {Slavin}, J.~D. 1994, \apjs, 95, 87

\bibitem[{{Werner} {et~al.}(2006){Werner}, {B{\"o}hringer}, {Kaastra}, {de
  Plaa}, {Simionescu}, \& {Vink}}]{werner2006}
{Werner}, N., {B{\"o}hringer}, H., {Kaastra}, J.~S., {et~al.} 2006, \aap, 459,  353

\bibitem[{{Werner} {et~al.}(2010){Werner}, {Simionescu}, {Million}, {Allen},
  {Nulsen}, {von der Linden}, {Hansen}, {B{\"o}hringer}, {Churazov}, {Fabian},
  {Forman}, {Jones}, {Sanders}, \& {Taylor}}]{2010MNRAS.407.2063W}
{Werner}, N., {Simionescu}, A., {Million}, E.~T., {et~al.} 2010, \mnras, 407,  2063

\end{thebibliography}

\clearpage

\begin{figure}
\epsscale{.80}
\plotone{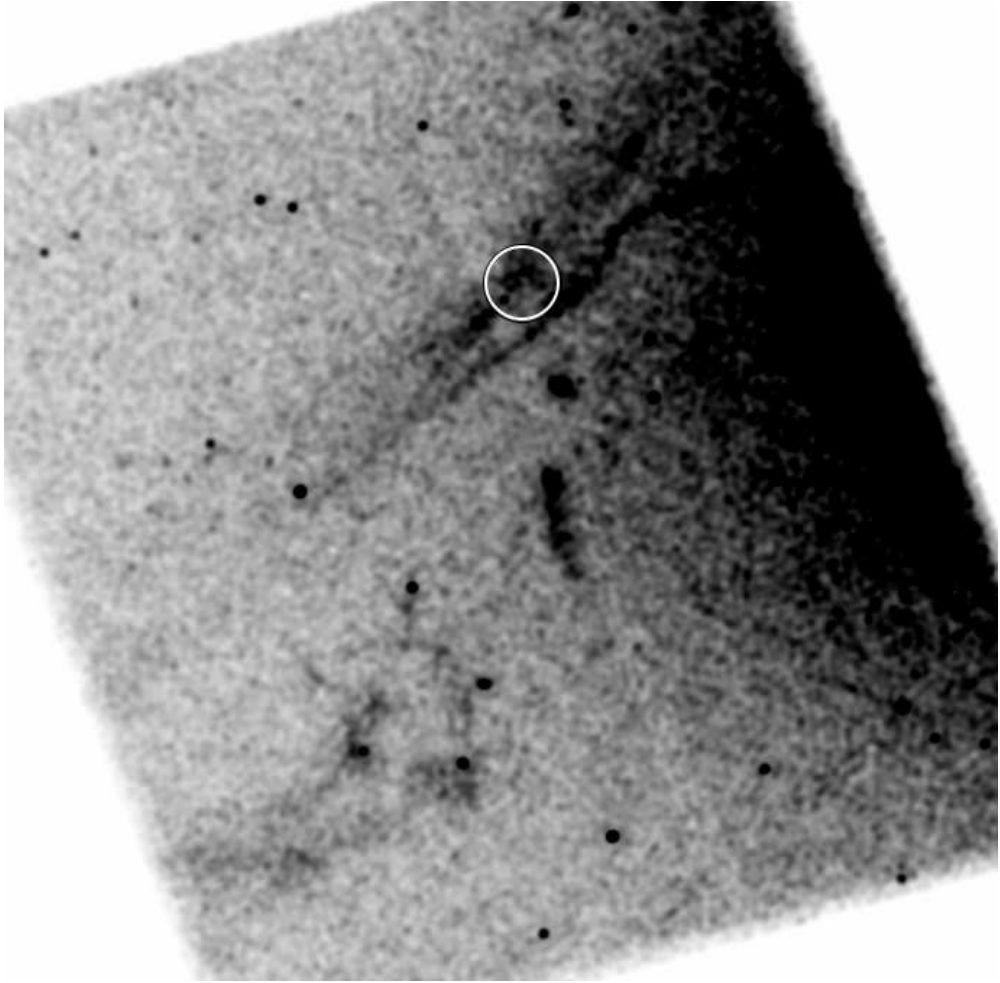}
\caption{Sum of four individual exposures of the M87 filament system taken with the Advanced Camera for Surveys Solar Blind Channel (ACS/SBC) using the 150~nm long-pass filter, which transmits a far-ultraviolet spectral band containing the 
C\,IV~1549~\AA\  emission line.\label{fig1}}
\end{figure}
\clearpage

\begin{figure}
\plotone{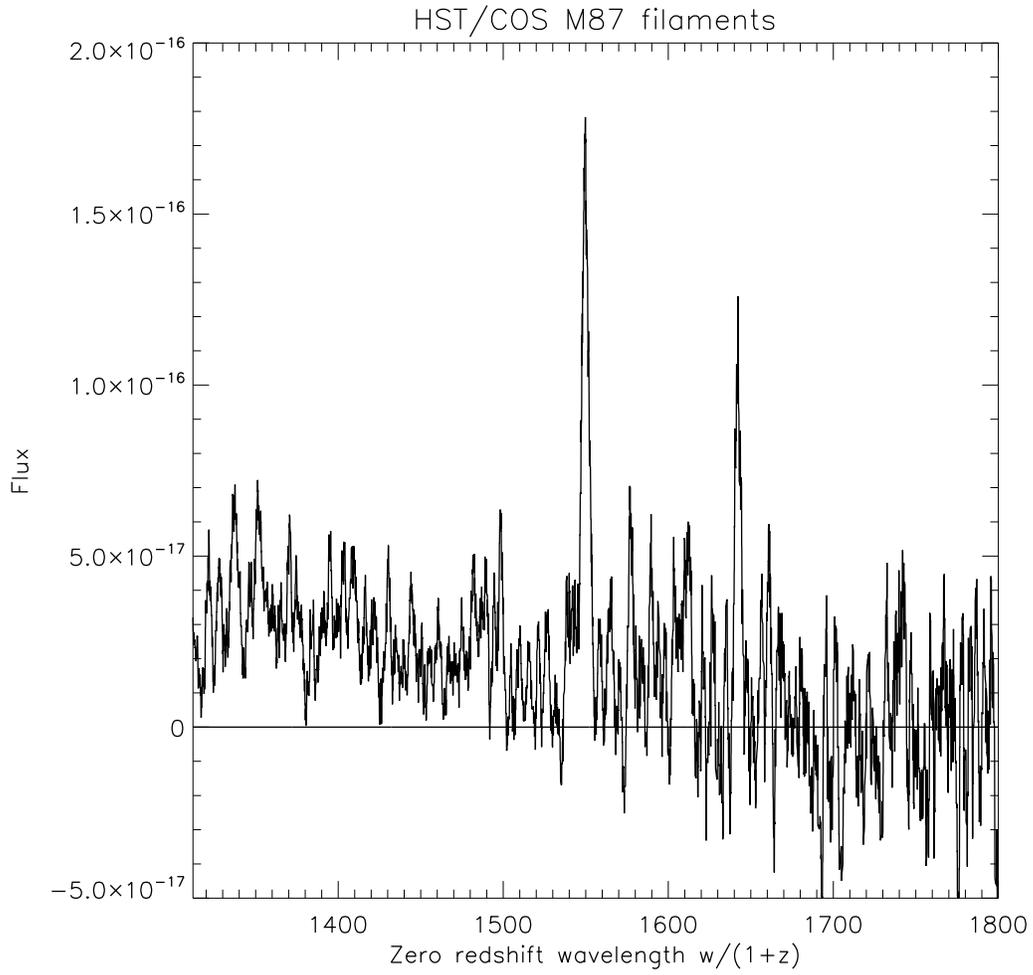}
\caption{Median spectrum of the four FUV/COS spectra, from two orbits of Hubble observations, smoothed with a bin size 2.56~\AA . The C\,IV~1549~\AA\  and 
He\,II~1640~\AA\  lines are clearly visible.\label{fig2}}
\end{figure}

\end{document}